\documentclass[reprint,superscriptaddress,preprintnumbers,nofootinbib,amsmath,amssymb,aps,prd,showkeys,showpacs]{revtex4-2}

\usepackage{graphicx}% Include figure files
\usepackage{dcolumn}% Align table columns on decimal point
\usepackage{bm}% bold math
\usepackage{array} % 导言区引入
\newcolumntype{M}[1]{>{\centering\arraybackslash}m{#1}}
\usepackage{endnotes} 
\usepackage{adjustbox}
\usepackage{changes}   % 在导言区

\usepackage{amssymb}    %Allows use math symbols
\usepackage{color}         %Allows {\color{red} Hello World} or \colorbox{blue}{Hello World}\add{the appendix part}.
\usepackage{graphicx}     %Allows\includegraphics{figure.pdf}
\usepackage{mathrsfs} %Allows \mathscr{HELLO}
\usepackage{hyperref} %Automatically links \label and \ref commands; Always load last
\hypersetup{colorlinks=true,linkcolor=blue,citecolor=blue}
\usepackage{ upgreek } % for uptau
\usepackage{color,xcolor,fancybox,epsf,rotating,colordvi}
\usepackage{booktabs}
\usepackage{multirow}

\usepackage{soul}
\setstcolor{red}
\usepackage[normalem]{ulem}  % [normalem]防止\emph被重定义
\usepackage{cancel}

\begin{document}

% \preprint{APS/123-QED}
\title{Gravitational waves in the complex-singlet-extended Manohar-Wise model}

\author{Benhong Mei}
\email[]{meibh99@henu.edu.cn}
\affiliation{School of Physics and Electronics, Henan University, Kaifeng 475004, China}

\author{Yabo Dong}
\email[]{dongyb@henu.edu.cn}
\affiliation{School of Physics and Electronics, Henan University, Kaifeng 475004, China}

\author{Jun Fan}
\email[]{fanjun@henu.edu.cn}
\affiliation{School of Physics and Electronics, Henan University, Kaifeng 475004, China}

\author{Jingya Zhu}
% \email[]{zhujy@henu.edu.cn} 
\email[Corresponding author:]{zhujy@henu.edu.cn} 
\affiliation{School of Physics and Electronics, Henan University, Kaifeng 475004, China}

% \collaboration{CLEO Collaboration}%\noaffiliation

\date{\today}% It is always \today, today,

\begin{abstract}

We investigate cosmological phase transitions and stochastic gravitational-wave signals in the Manohar–Wise model extended by a complex scalar singlet. Spontaneous breaking of a discrete $\mathbb{Z}_3$ symmetry produces domain walls, whose subsequent annihilation is driven by a small explicit symmetry-breaking bias. Subject to theoretical and experimental constraints, we study two thermal histories: a one-step strong first-order electroweak phase transition satisfying the baryon-number preservation criterion, and a two-step history whose first transition is a first-order transition along the singlet direction, leaving electroweak symmetry unbroken. We estimate the gravitational-wave contributions from sound waves, magnetohydrodynamic turbulence, and domain-wall annihilation. Within the adopted approximations, the phase-transition signals of the selected benchmarks peak at approximately $10^{-3}$–$10^{-2}\,\mathrm{Hz}$ and overlap the projected sensitivity regions of space-based observatories such as LISA, TianQin, and BBO. Choosing the vacuum-energy bias to fix the domain-wall peak at approximately $2\times10^{-9}\,\mathrm{Hz}$, we find that the larger domain-wall tensions of the two-step benchmarks yield signals reaching the projected SKA sensitivity, whereas the one-step sample remains below the pulsar-timing sensitivities considered. These results illustrate the complementarity of millihertz and nanohertz gravitational-wave observations in probing phase transitions and discrete-symmetry breaking in this model.
% 对MW模型进行复单态扩充，研究其引力波特性。
% 引力波主要来自于相变的SW，tub和筹壁，对比一步与二步相变。
% SW峰值在哪个实验观测频段，筹壁在哪个频段
% 哪些可以解释重子数不对称
% 哪些在纳Hz频段可以被看到
\end{abstract}

\maketitle
\newpage

% \tableofcontents
% \newpage

%%%

\section{Introduction}
\label{sec:intro}

The first direct detection of gravitational waves (GWs) from a binary black hole merger by the Laser Interferometer Gravitational-Wave Observatory (LIGO) opened a new window on the Universe~\cite{LIGOScientific:2016aoc}.
Beyond signals from black hole and neutron star mergers, potential cosmological GW sources have attracted considerable interest, including cosmic strings~\cite{Ellis:2023cs, Bian:2023gw, Wang:2023cs, Lazarides:2023dw, Eichhorn:2024gw, Chowdhury:2023de, Antusch:2023so10, Yamada:2023dw, Ge:2024gw, Basilakos:2024gw, Avgoustidis:2025svu}, domain walls~\cite{Kitajima:2024dw, Blasi:2023axion, Gouttenoire:2024dw, Lu:2024clockwork, Babichev:2023dw, Gelmini:2024axion, Guo:2024axion, Zhang:2023dw, Du:2023susy, Li:2025axion, Li:2023gil, Lu:2024bh,Xu:2025vmy}, cosmological phase transitions and related dynamics~\cite{Fujikura:2023pt, Franciolini:2024qcd, Bringmann:2023pt, Addazi:2024pt, Bai:2023dwqcd, Han:2024sidm, Zu:2024mirror, Ghosh:2024pt, Li:2023collider, DiBari:2024majoron, Cruz:2023eden, Gouttenoire:2023ptbh, Ahmadvand:2023qcdaxion, Salvio:2023sc, Athron:2024scpt, Jiang:2024dm, He:2025holographic, Chao:2024dp, Goncalves:2025uwh, Costa:2025csj, Salvio:2024ptpbh}, and other early-Universe mechanisms~\cite{Vagnozzi:2023inf, Frosina:2023inf, Liu:2024ng, Unal:2024axioninf, Bari:2024scalar, Das:2023quint, Jiang:2024infgw, Gorji:2023pti, An:2024infpt, Cang:2025hfgw, Liu:2024eos}.
On the observational side, pulsar timing arrays (PTAs), including the North American Nanohertz Observatory for Gravitational Waves (NANOGrav)~\cite{Afzal:2023nanograv, Agazie:2023nanograv1, Agazie:2023nanograv2}, the European Pulsar Timing Array (EPTA)~\cite{Antoniadis:2023ept1, Antoniadis:2023ept2}, the Parkes Pulsar Timing Array (PPTA)~\cite{Reardon:2023ppta}, and the Chinese Pulsar Timing Array (CPTA)~\cite{Xu:2023cpta}, probe the nanohertz GW background and constrain its possible origins.
The future space-based Laser Interferometer Space Antenna (LISA) will probe GWs in the millihertz band, providing a complementary observational window~\cite{LISA:2017pwj}.

GWs generated by first-order phase transitions (FOPTs) provide a probe of particle physics in the early Universe.
However, for the observed Higgs boson mass, the electroweak transition in the Standard Model (SM) is a smooth crossover~\cite{DOnofrio:2014rug} rather than a first-order transition, and therefore does not produce GWs through bubble nucleation and expansion.
In theories beyond the Standard Model (BSM), extending the scalar sector can give rise to a strong first-order electroweak phase transition.
For example, an additional scalar singlet coupled to the Higgs doublet can enable such a transition in a singlet extension of the SM~\cite{Xiao:2023dbb, Papadopoulos:2026tmu, Pham:2026bip, Liang:2025zku, Ghosh:2025rbt, Ghosh:2022fzp, Ellis:2022lft, Cho:2022our, Wang:2022lxn, Han:2020ekm, Chen:2020wvu, Barman:2019oda, Shajiee:2018jdq, Imtiaz:2018dfn, Matsui:2017ggm, Kang:2017mkl, Vaskonen:2016yiu, Balazs:2016tbi, Balazs:2016tbi, Hashino:2016xoj}.
Scalar extensions can also accommodate spontaneously broken discrete symmetries, which can lead to the formation of domain walls~\cite{Kibble:1976sj}.
In the next-to-two-Higgs-doublet model (N2HDM), for instance, spontaneous breaking of the discrete $\mathbb{Z}_2$ symmetry associated with the real singlet scalar can lead to domain-wall formation, and GWs from the subsequent annihilation of these walls provide a probe of this scenario~\cite{Xu:2025vmy, Gustavsson:2026cps, Chaudhuri:2026ppk, Biermann:2024oyy, Sassi:2024cyb, Chaudhuri:2024vrd}.

The Manohar--Wise (MW) model extends the SM by introducing a scalar field transforming as $(8,2)_{1/2}$ under the gauge group $SU(3)_C \times SU(2)_L \times U(1)_Y$~\cite{Manohar:2006ga}.
The model implements minimal flavor violation (MFV), which constrains the flavor structure of the new scalar couplings and suppresses potentially dangerous flavor-changing neutral currents (FCNCs).
More generally, color-octet scalars appear in a variety of theoretical frameworks, including grand unified theories~\cite{Bertolini:2013vta}, supersymmetric models~\cite{Plehn:2008ae, Choi:2008ub}, top-coloron models~\cite{Chivukula:2013kw}, and models with extra dimensions~\cite{Dobrescu:2007xf}.
The new scalars can modify Higgs production rates and give rise to additional collider signatures through their direct production and subsequent decays~\cite{Manohar:2006ga, Dobrescu:2007yp}.
To incorporate domain-wall formation into this framework, we extend the scalar sector to include a discrete symmetry that is spontaneously broken in the vacuum.

In this work, we extend the Manohar--Wise model~\cite{Manohar:2006ga, Cao:2013wqa} by introducing a complex gauge-singlet scalar field $S$.
The cubic term $S^3 + S^{\dagger 3}$ in the scalar potential explicitly breaks the global $U(1)$ symmetry of the singlet sector down to a discrete $\mathbb{Z}_3$ symmetry.
A nonzero singlet vacuum expectation value (VEV) spontaneously breaks this discrete symmetry, allowing domain walls to form between distinct degenerate vacua.
A small explicit breaking of the $\mathbb{Z}_3$ symmetry introduces an energy bias between these vacua and drives the subsequent annihilation of the domain walls.
Within this framework, we investigate one-step and two-step phase-transition histories, including a one-step strong first-order electroweak phase transition, and calculate the stochastic GW spectra sourced by sound waves, magnetohydrodynamic (MHD) turbulence, and domain-wall annihilation.
We assess the prospects for detecting the phase-transition signals with future space-based interferometers such as LISA and the domain-wall signals with pulsar timing observations using facilities such as the Square Kilometre Array (SKA).

The remaining part of this work is organized as follows: Section~\ref{sec:model} introduces the complex-singlet-extended Manohar-Wise model. In Section~\ref{sec:GW}, these GW signals from the phase transition and domain wall are described in detail. In Section~\ref {parameter_scan}, we scan the parameter space with some constraints and discuss the scanning results and GW signals.
Finally, the conclusion is summarized in Section~\ref{sec:conclusion}.
%%%%%%%%%%%%%%%%%%%%%%%%%%%%%%%%%%%%%%%%%%%%%%%%%%%%%%%%%%%%%%%%%%%%%%%%%%%%%%%%%%%%%%%%%%%%%%
\section{\label{sec:model}The complex-singlet-extended Manohar-Wise model (MWS)} 

\subsection{The model}
We extend the Manohar--Wise model by introducing a complex scalar singlet.
The scalar sector consists of the SM Higgs doublet $H \sim (1,2)_{1/2}$, a complex gauge singlet $S \sim (1,1)_0$, and a color-octet scalar doublet $\Phi \sim (8,2)_{1/2}$~\cite{Manohar:2006ga, Cao:2013wqa}, where the representations refer to $SU(3)_C \times SU(2)_L \times U(1)_Y$.
The fields are parametrized as
\begin{equation}
    H =
    \begin{pmatrix}
        G^+ \\
        \dfrac{h+iG^0}{\sqrt{2}}
    \end{pmatrix},
    \quad
    S = \frac{s+i\chi}{\sqrt{2}},
    \quad
    \Phi^A =
    \begin{pmatrix}
        \phi_+^A \\
        \dfrac{\phi_R^A+i\phi_I^A}{\sqrt{2}}
    \end{pmatrix}.
\end{equation}
Here, $A=1,\ldots,8$ is an adjoint color index.
The fields $\phi_+^A$ are the charged components of the color-octet doublet, while $\phi_R^A$ and $\phi_I^A$ denote its neutral real and imaginary components, respectively.
We choose a vacuum characterized by
\begin{equation}
    \langle h\rangle = v,
    \qquad
    \langle s\rangle = v_s,
    \qquad
    \langle\chi\rangle = 0,
\end{equation}
so that $\langle H\rangle=(0,v/\sqrt{2})^T$ and
$\langle S\rangle=v_s/\sqrt{2}$.

Under the discrete $\mathbb{Z}_3$ symmetry, the singlet transforms as
$S \to e^{2\pi i/3}S$, while $H$ and $\Phi$ remain unchanged.
The $\mathbb{Z}_3$-symmetric scalar potential is given by
\begin{equation}
\begin{aligned}
V ={}&
-\mu_H^2 H^\dagger H
+\lambda_H(H^\dagger H)^2
-\mu_s^2 |S|^2
+\lambda_s |S|^4
\\
&+\lambda_{sh}(H^\dagger H)|S|^2
+\frac{u_3}{2}\left(S^3+S^{\dagger 3}\right)
\\
&+2m_s^2\operatorname{Tr}
  \left(\Phi^{\dagger i}\Phi_i\right)
+\lambda |S|^2\operatorname{Tr}
  \left(\Phi^{\dagger i}\Phi_i\right)
\\
&+\lambda_1 H^{\dagger i}H_i
  \operatorname{Tr}\left(\Phi^{\dagger j}\Phi_j\right)
+\lambda_2 H^{\dagger i}H_j
  \operatorname{Tr}\left(\Phi^{\dagger j}\Phi_i\right)
\\
&+\Bigl[
  \lambda_3 H^{\dagger i}H^{\dagger j}
  \operatorname{Tr}\left(\Phi_i\Phi_j\right)
\\
&\qquad
  +\lambda_4 H^{\dagger i}
  \operatorname{Tr}\left(\Phi^{\dagger j}\Phi_j\Phi_i\right)
\\
&\qquad
  +\lambda_5 H^{\dagger i}
  \operatorname{Tr}\left(\Phi^{\dagger j}\Phi_i\Phi_j\right)
  +\mathrm{h.c.}
  \Bigr]
\\
&+\lambda_6\operatorname{Tr}
  \left(\Phi^{\dagger i}\Phi_i\Phi^{\dagger j}\Phi_j\right)
\\
&+\lambda_7\operatorname{Tr}
  \left(\Phi^{\dagger i}\Phi_j\Phi^{\dagger j}\Phi_i\right)
\\
&+\lambda_8
  \operatorname{Tr}\left(\Phi^{\dagger i}\Phi_i\right)
  \operatorname{Tr}\left(\Phi^{\dagger j}\Phi_j\right)
\\
&+\lambda_9
  \operatorname{Tr}\left(\Phi^{\dagger i}\Phi_j\right)
  \operatorname{Tr}\left(\Phi^{\dagger j}\Phi_i\right)
\\
&+\lambda_{10}
  \operatorname{Tr}\left(\Phi_i\Phi_j\right)
  \operatorname{Tr}\left(\Phi^{\dagger i}\Phi^{\dagger j}\right)
\\
&+\lambda_{11}\operatorname{Tr}
  \left(\Phi_i\Phi_j\Phi^{\dagger j}\Phi^{\dagger i}\right).
\end{aligned}
\end{equation}
Here, $i,j=1,2$ are $SU(2)_L$ indices, and repeated indices are summed.
The traces are taken over color indices, with
$\Phi_i=\Phi_i^A T^A$, where $T^A$ are the generators of
$SU(3)_C$ in the fundamental representation, normalized as
$\operatorname{Tr}(T^A T^B)=\delta^{AB}/2$.
The coefficient $u_3$ is taken to be real.
We consider a color-preserving vacuum with
$\langle\Phi_i^A\rangle=0$.

The cubic term explicitly breaks the global $U(1)$ symmetry of
the singlet sector down to a discrete $\mathbb{Z}_3$ symmetry.
The terms proportional to $\lambda_4,\ldots,\lambda_{11}$ vanish
on the color-preserving background $\Phi=0$.
To simplify the parameter space, we set these couplings to zero
in the present analysis.
Taking $\lambda_3$ to be real and setting $G^\pm=G^0=0$,
we write the tree-level potential in terms of the scalar
components as
\begin{equation}
\begin{aligned}
V_0 ={}&
-\frac{\mu_H^2}{2}h^2
+\frac{\lambda_H}{4}h^4
-\frac{\mu_s^2}{2}(s^2+\chi^2)
+\frac{\lambda_s}{4}(s^2+\chi^2)^2
\\
&+\frac{\lambda_{sh}}{4}h^2(s^2+\chi^2)
+\frac{u_3}{2\sqrt{2}}\left(s^3-3s\chi^2\right)
\\
&+\left[
m_s^2+\frac{\lambda_1}{4}h^2
+\frac{\lambda}{4}(s^2+\chi^2)
\right]\phi_+^A\phi_-^A
\\
&+\frac{1}{2}\left[
m_s^2+\frac{\lambda_1+\lambda_2+2\lambda_3}{4}h^2
+\frac{\lambda}{4}(s^2+\chi^2)
\right](\phi_R^A)^2
\\
&+\frac{1}{2}\left[
m_s^2+\frac{\lambda_1+\lambda_2-2\lambda_3}{4}h^2
+\frac{\lambda}{4}(s^2+\chi^2)
\right](\phi_I^A)^2.
\end{aligned}
\end{equation}
Here, $\phi_-^A=(\phi_+^A)^*$, and the repeated color index
$A$ is summed.
The tree-level potential along the color-preserving background
is obtained by setting
$\phi_\pm^A=\phi_R^A=\phi_I^A=0$.
    
The stationarity conditions at the color-preserving vacuum
$(h,s,\chi)=(v,v_s,0)$ are
\begin{equation}
\left.\frac{\partial V_0}{\partial h}\right|_{(v,v_s,0)}=0,
\qquad
\left.\frac{\partial V_0}{\partial s}\right|_{(v,v_s,0)}=0,
\end{equation}
while the condition
$\left.\partial V_0/\partial\chi\right|_{(v,v_s,0)}=0$
is automatically satisfied.
For nonzero $v$ and $v_s$, these conditions yield
\begin{equation}
\begin{aligned}
\mu_H^2 &=
\lambda_H v^2+\frac{\lambda_{sh}}{2}v_s^2,
\\
\mu_s^2 &=
\lambda_s v_s^2+\frac{\lambda_{sh}}{2}v^2
+\frac{3u_3v_s}{2\sqrt{2}}.
\end{aligned}
\end{equation}
Defining the fluctuations around the vacuum as
$\delta h=h-v$ and $\delta s=s-v_s$, we obtain the
tree-level mass-squared matrix in the $(\delta h,\delta s)$
basis:
\begin{equation}
M^2=
\begin{pmatrix}
2\lambda_H v^2
&
\lambda_{sh}vv_s
\\
\lambda_{sh}vv_s
&
2\lambda_s v_s^2+\dfrac{3u_3v_s}{2\sqrt{2}}
\end{pmatrix}.
\end{equation}

At zero temperature, local stability in the $(\delta h,\delta s)$ sector requires the corresponding mass-squared matrix to be positive definite. For $v>0$ and $v_s>0$, this is equivalent to
\begin{equation}
    \lambda_H>0, \quad 8v_s\lambda_H\lambda_s-2v_s\lambda_{sh}^2+3\sqrt{2}\lambda_Hu_3>0.
\end{equation}
These conditions ensure local stability only in the
$(\delta h,\delta s)$ sector.
Stability in the $\chi$ and color-octet directions must
also be checked, and establishing the vacuum as the
global minimum requires a separate analysis.

The scalar mass eigenstates $h_1$ and $h_2$ are obtained
by an orthogonal rotation of the fluctuations
$\delta h$ and $\delta s$:
\begin{equation}
\begin{pmatrix}
h_1\\
h_2
\end{pmatrix}
=
R(\theta)
\begin{pmatrix}
\delta h\\
\delta s
\end{pmatrix},
\quad
R(\theta)=
\begin{pmatrix}
\cos\theta & \sin\theta\\
-\sin\theta & \cos\theta
\end{pmatrix}.
\end{equation}
The mass-squared matrix is diagonalized according to
\begin{equation}
R(\theta)M^2R^T(\theta)
=
\begin{pmatrix}
m_1^2 & 0\\
0 & m_2^2
\end{pmatrix},
\label{eq:q1}
\end{equation}
where $m_1$ and $m_2$ are the masses of $h_1$ and $h_2$,
respectively.
Using Eq.~\eqref{eq:q1}, we obtain
\begin{equation}
\begin{aligned}
m_1^2 ={}&
2\lambda_H v^2\cos^2\theta
+\left(
2\lambda_s v_s^2+\frac{3u_3v_s}{2\sqrt{2}}
\right)\sin^2\theta
\\
&+2\lambda_{sh}vv_s\sin\theta\cos\theta,
\\
m_2^2 ={}&
2\lambda_H v^2\sin^2\theta
+\left(
2\lambda_s v_s^2+\frac{3u_3v_s}{2\sqrt{2}}
\right)\cos^2\theta
\\
&-2\lambda_{sh}vv_s\sin\theta\cos\theta.
\end{aligned}
\end{equation}
The mixing angle satisfies
\begin{equation}
\tan 2\theta
=
\frac{\lambda_{sh}vv_s}
{\lambda_Hv^2-\lambda_sv_s^2
-\dfrac{3u_3v_s}{4\sqrt{2}}}.
\end{equation}
We identify $h_1$ with the observed SM-like Higgs boson
and fix $m_1=125\,\mathrm{GeV}$, while taking $m_2>m_1$.
The cubic term explicitly breaks the global $U(1)$
symmetry of the singlet sector down to $\mathbb{Z}_3$,
giving a mass to the would-be Goldstone mode $\chi$.
Using the stationarity conditions, we obtain its tree-level
mass squared as
\begin{equation}
m_\chi^2
=
\left.
\frac{\partial^2 V_0}{\partial\chi^2}
\right|_{(v,v_s,0)}
=
-\frac{9u_3v_s}{2\sqrt{2}}.
\end{equation}
Local stability in the $\chi$ direction requires
$m_\chi^2>0$, implying $u_3v_s<0$.
With the convention $v_s>0$, this requires $u_3<0$.

With $\lambda_4=\cdots=\lambda_{11}=0$, the conditions
$m_s^2>0$, $\lambda>0$, $\lambda_1>0$, and
$\lambda_1+\lambda_2\pm2\lambda_3>0$ ensure that, for
any fixed Higgs and singlet background, the tree-level
potential is minimized at $\Phi=0$.
The search for the global minimum can therefore be
restricted to the color-preserving Higgs--singlet
subspace.
Within this subspace, we impose the tree-level vacuum
constraint of Ref.~\cite{Zhou:2020ojf},
\begin{equation}
m_\chi^2
<
\frac{9m_1^2m_2^2}
{m_1^2\cos^2\theta+m_2^2\sin^2\theta}.
\end{equation}

At the vacuum $(h,s,\chi,\Phi)=(v,v_s,0,0)$, the
tree-level squared masses of the color-octet scalars are
\begin{equation}
\begin{aligned}
M_\pm^2 &=
m_s^2+\frac{\lambda_1}{4}v^2
+\frac{\lambda}{4}v_s^2,
\\
M_R^2 &=
m_s^2+\frac{\lambda_1+\lambda_2+2\lambda_3}{4}v^2
+\frac{\lambda}{4}v_s^2,
\\
M_I^2 &=
m_s^2+\frac{\lambda_1+\lambda_2-2\lambda_3}{4}v^2
+\frac{\lambda}{4}v_s^2,
\end{aligned}
\end{equation}
where $M_\pm$, $M_R$, and $M_I$ denote the masses of
$\phi_\pm^A$, $\phi_R^A$, and $\phi_I^A$, respectively.
The masses are independent of the color index $A$.

To suppress the color-octet contribution to the
electroweak $T$ parameter, we consider regions in which
the charged scalar is approximately degenerate with
one of the neutral scalars~\cite{Burgess:2009wm}.
Specifically, we take
\begin{equation}
    \begin{aligned}
        &\lambda_2 \simeq 2\lambda_3, \quad M_\pm \simeq M_I,\\
    \end{aligned}
\end{equation}
or
\begin{equation}
    \begin{aligned}
    &\lambda_2 \simeq -2\lambda_3, \quad M_\pm \simeq M_R.
    \end{aligned}
\end{equation}

The scalar-potential parameters can be expressed in terms
of the tree-level masses $m_1$, $m_2$, $m_\chi$,
$M_\pm$, $M_R$, and $M_I$, the VEVs $v$ and $v_s$,
the mixing angle $\theta$, and the independent parameters
$m_s$ and $\lambda$:
\begin{equation}
\begin{aligned}
\lambda_H &=
\frac{m_1^2\cos^2\theta+m_2^2\sin^2\theta}{2v^2},
\\[6pt]
\lambda_s &=
\frac{3m_1^2\sin^2\theta+3m_2^2\cos^2\theta+m_\chi^2}
{6v_s^2},
\\[6pt]
\lambda_{sh} &=
\frac{(m_1^2-m_2^2)\sin 2\theta}{2vv_s},
\\[6pt]
u_3 &=
-\frac{2\sqrt{2}\,m_\chi^2}{9v_s},
\\[6pt]
\lambda_1 &=
\frac{4M_\pm^2-4m_s^2-\lambda v_s^2}{v^2},
\\[6pt]
\lambda_2 &=
\frac{2(M_R^2+M_I^2-2M_\pm^2)}{v^2},
\\[6pt]
\lambda_3 &=
\frac{M_R^2-M_I^2}{v^2}.
\end{aligned}
\end{equation}

\subsection{Finite temperature potential}
The finite temperature potential utilizes the gauge-invariant approach \cite{Bian:2019bsn, Alves:2018jsw}, which was adopted for the study of phase transition behavior in the complex singlet model, as given by \cite{Zhou:2020ojf},
 \begin{equation}
        \begin{aligned}
            V_T = &(c_{hT}-\frac{1}{2}\mu_H^2)h^2 +\frac{1}{4}\lambda_H h^4 + (c_{sT}-\frac{1}{2}\mu_s^2)s^2 \\
             &+ \frac{1}{4}\lambda_s s^4 + \frac{1}{4}\lambda_{sh}h^2s^2 + \frac{\sqrt{2}}{4}u_3s^3 ,
        \end{aligned}
 \end{equation}
and the finite temperature correction term is
 \begin{equation}
  \begin{aligned}
   c_{hT} =& \frac{T^2}{48}(\frac{9}{2}g^2 + \frac{3}{2}g^{\prime 2}+6y_t^2 + 12\lambda_H + 2\lambda_{sh} \\
            &+ 16\lambda_1 + 8\lambda_2)=a_h T^2 ,\\
   c_{sT} =& \frac{T^2}{24}(4\lambda_s + 2\lambda_{sh} + 8\lambda)=a_s T^2\ .
  \end{aligned}
 \end{equation}
Compared with the complex extended SM, there are several more parameters in the correction terms. 
%%%%%%%%%%%%%%%%%%%%%%%%%%%%%%%%%%%%%%%%%%%%%%%%%%%%%%%%%%%%%%%%%%%%%%%%%%%%%%%%%%%%%%%%%%%%%%
\section{\label{sec:GW}Gravitational waves} 

\subsection{The strong first-order phase transition}
For a first-order phase transition, as the temperature falls below the critical temperature $T_c$, the low-temperature phase becomes energetically favored, while the high-temperature phase remains metastable.
The false vacuum can decay through thermal nucleation of bubbles of the lower-energy phase.
The bubble nucleation rate per unit volume and time is given by
\begin{equation}
\Gamma(T)\simeq A(T)\exp\left[-\frac{S_3(T)}{T}\right],
\end{equation}
where $A(T)$ is the prefactor and $S_3(T)$ is the three-dimensional Euclidean action evaluated on the $O(3)$-symmetric bounce solution:
\begin{equation}
    \begin{aligned}
        S_3(T)=4\pi\int_0^\infty dr\,r^2
            \Biggl[
            &\frac{1}{2}\left(\frac{dh}{dr}\right)^2
            +\frac{1}{2}\left(\frac{ds}{dr}\right)^2 \\
            &+V_T(h,s;T)-V_T(h_f,s_f;T) 
            \Biggr].
    \end{aligned}
\end{equation}
Here, $r$ is the radial coordinate, and $(h_f,s_f)$ denotes the false vacuum at temperature $T$.

Varying $S_3(T)$ with respect to $h(r)$ and $s(r)$ gives the coupled bounce equations,
\begin{equation}
\begin{aligned}
\frac{d^2h}{dr^2}+\frac{2}{r}\frac{dh}{dr}
&=\frac{\partial V_T(h,s;T)}{\partial h},
\\
\frac{d^2s}{dr^2}+\frac{2}{r}\frac{ds}{dr}
&=\frac{\partial V_T(h,s;T)}{\partial s}.
\end{aligned}
\end{equation}
The bounce profiles satisfy the boundary conditions
\begin{equation}
\begin{aligned}
\left.\frac{dh}{dr}\right|_{r=0}
&=\left.\frac{ds}{dr}\right|_{r=0}=0,
\\
\lim_{r\to\infty}h(r)&=h_f(T),
\\
\lim_{r\to\infty}s(r)&=s_f(T),
\end{aligned}
\end{equation}
which ensure regularity at the bubble center and approach to the false vacuum at spatial infinity.

The nucleation temperature $T_n$ is conventionally defined as the temperature at which the expected number of nucleated bubbles per Hubble volume becomes of order unity. Assuming adiabatic cooling with slowly varying effective entropy degrees of freedom, this criterion can be approximated by~\cite{Xiao:2023dbb}
\begin{equation}
\int_{T_n}^{T_c}\frac{dT}{T}\,
\frac{\Gamma(T)}{H^4(T)}
\simeq 1 \ .
\end{equation}
Here, $T_c$ is the critical temperature of the transition, and
\begin{equation}
H(T)=\sqrt{\frac{8\pi G}{3}\rho(T)}
\end{equation}
is the Hubble expansion rate, with $G$ denoting Newton's gravitational constant and $\rho(T)$ the total energy density of the Universe. For an electroweak-scale transition in a radiation-dominated Universe, a commonly used estimate is
\begin{equation}
    \frac{S_3(T_n)}{T_n}\simeq 140 \ .
\end{equation}
This estimate characterizes the onset of appreciable bubble nucleation and does not by itself guarantee completion of the phase transition.
To select the parameter points studied in this work, we require $v_n/T_n>1$ for the one-step sample and $v_{sn}/T_n>1$ for the two-step sample, where $v_n$ and $v_{sn}$ denote the values of $h$ and $s$, respectively, in the low-temperature phase at the nucleation temperature $T_n$ of the transition under consideration. The first ratio is a commonly used approximate measure of electroweak phase-transition strength. The second is adopted here as a selection criterion for the singlet direction and does not by itself imply a strong first-order electroweak phase transition or suppression of baryon-number washout.

For scenario A, we assess the preservation of a pre-existing baryon asymmetry against washout by electroweak sphaleron processes using the baryon-number preservation criterion (BNPC)~\cite{Zhou:2020ojf, Zhou:2020irf, Zhou:2019uzq},
\begin{equation}
\begin{aligned}
PT_{\mathrm{sph}}\equiv{}&
\frac{E_{\mathrm{sph}}(T_n)}{T_n}
-7\ln\frac{v(T_n)}{T_n}
+\ln\frac{T_n}{100\,\mathrm{GeV}},
\\
PT_{\mathrm{sph}}>{}&B_{\mathrm{sph}},
\quad
B_{\mathrm{sph}}\in[35.9,42.8].
\end{aligned}
\end{equation}
Here, $v(T_n)$ is the Higgs-doublet background value in the broken phase, and $E_{\mathrm{sph}}(T_n)$ is the sphaleron energy at the nucleation temperature. We evaluate the sphaleron energy following the methods described in Refs.~\cite{Zhou:2020ojf, Zhou:2020irf, Zhou:2019uzq}.

The range of $B_{\mathrm{sph}}$ mainly reflects the uncertainty in the sphaleron fluctuation determinant, $\kappa\sim10^{-4}$--$10^{-1}$~\cite{Dine:1991ck, Zhou:2020ojf}. Satisfying this criterion indicates suppression of baryon-number washout, but does not by itself establish successful electroweak baryogenesis.

\subsection{Phase-transition patterns}

\begin{figure}[htbp]
        \centering
        \includegraphics[width=0.5\textwidth]{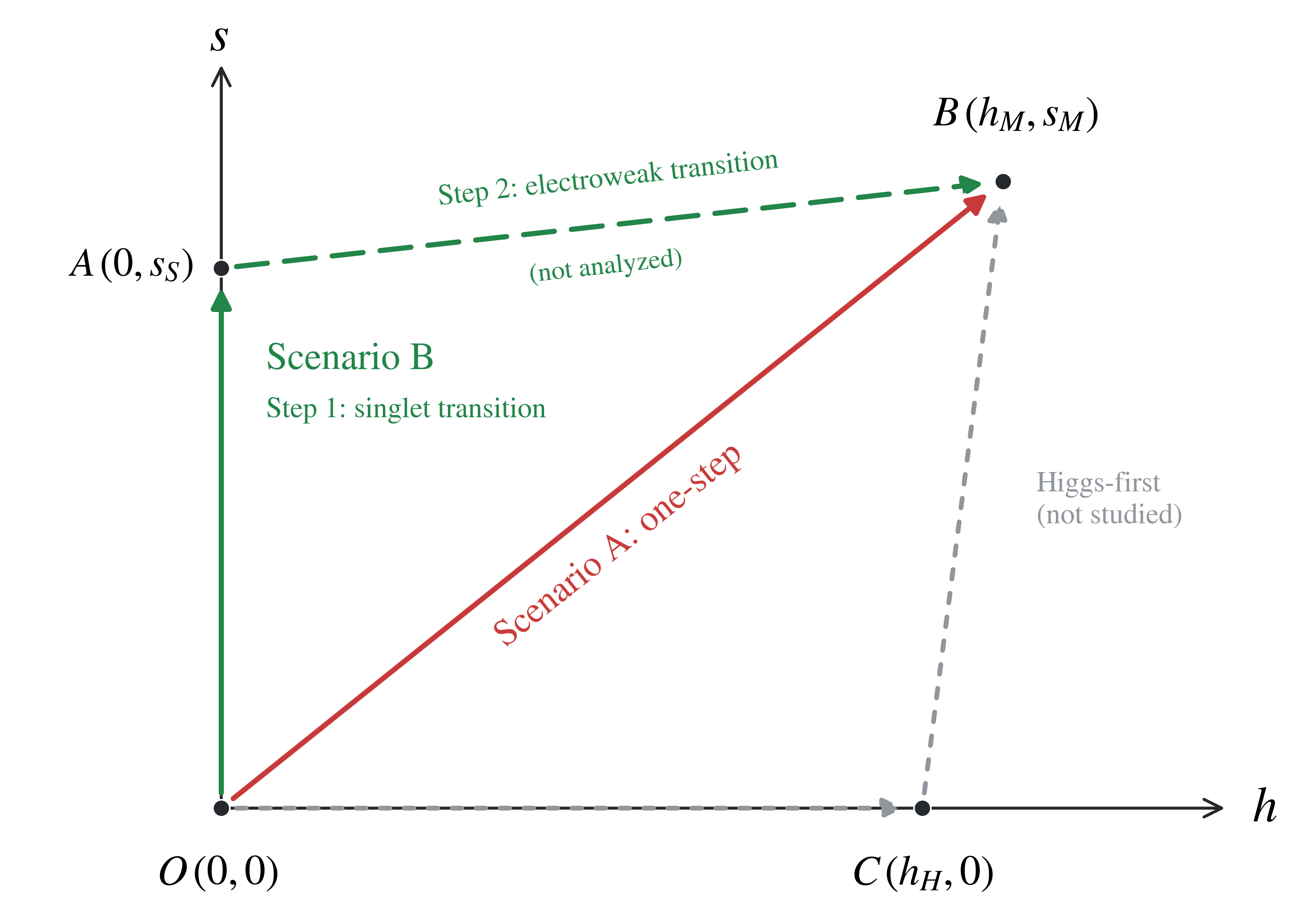}
        \caption{Schematic diagram of the possible phase-transition pathways.}
        \label{fig:my_label}
\end{figure}

Restricting the background fields to the $(h,s)$ subspace, we consider different thermal histories according to the sequence of electroweak and $\mathbb{Z}_3$ symmetry breaking. In a one-step transition, both symmetries are broken in a single transition from the symmetric phase to a phase with nonzero Higgs and singlet background values. In a two-step transition, symmetry breaking proceeds through an intermediate phase in which only one of the two fields has a nonzero background value, followed by a transition to the phase in which both are nonzero. The two stages occur at different temperatures and need not both be first order. The possible transition paths are illustrated in Figure~\ref{fig:my_label}.

For a first-order transition from the symmetric phase at $(h,s)=(0,0)$, the critical temperature $T_c$ is defined by the degeneracy of the symmetric minimum and a distinct minimum at $(h_c,s_c)$. The critical temperature and the corresponding background field values satisfy
\begin{equation}
\begin{aligned}
V_T(0,0;T_c)&=V_T(h_c,s_c;T_c),
\\
\left.\frac{\partial V_T}{\partial h}\right|_{(h_c,s_c;T_c)}&=0,
\\
\left.\frac{\partial V_T}{\partial s}\right|_{(h_c,s_c;T_c)}&=0.
\end{aligned}
\label{eq:q29}
\end{equation}
For a transition from an intermediate phase, the first equality must instead involve the potential evaluated at the corresponding high-temperature minimum.

Both degenerate stationary points must be local minima. At $(h_c,s_c)$, local stability in the $(h,s)$ subspace requires
\begin{equation}
M_1>0,
\quad
M_1P_1-N_1^2>0,
\end{equation}
where
\begin{equation}
\begin{aligned}
M_1&=\left.\frac{\partial^2V_T}{\partial h^2}\right|_{(h_c,s_c;T_c)},
\\
P_1&=\left.\frac{\partial^2V_T}{\partial s^2}\right|_{(h_c,s_c;T_c)},
\\
N_1&=\left.\frac{\partial^2V_T}{\partial s\,\partial h}\right|_{(h_c,s_c;T_c)}.
\end{aligned}
\label{eq:q30}
\end{equation}
At the symmetric minimum, the corresponding conditions are
\begin{equation}
-\mu_H^2+2a_hT_c^2>0,
\quad
-\mu_s^2+2a_sT_c^2>0.
\end{equation}

For a transition from the origin to a minimum with $h_c>0$ and $s_c>0$, the stationarity and degeneracy conditions in Eq.~\eqref{eq:q29} give
\begin{equation}
\begin{aligned}
\lambda_Hh_c^2+\frac{\lambda_{sh}}{2}s_c^2
&=\mu_H^2-2a_hT_c^2,
\\
\frac{\lambda_{sh}}{2}h_c^2+\lambda_ss_c^2
+\frac{3u_3}{2\sqrt{2}}s_c
&=\mu_s^2-2a_sT_c^2,
\\
\lambda_Hh_c^4+\lambda_{sh}h_c^2s_c^2
+\lambda_ss_c^4+\frac{u_3}{\sqrt{2}}s_c^3
&=0.
\end{aligned}
\label{eq:q31}
\end{equation}

For a transition from the origin to a singlet minimum with $h_c=0$ and $s_c>0$, the same conditions yield
\begin{equation}
    \begin{aligned}
        s_c   &=-\frac{u_3}{\sqrt{2}\lambda_s},\\
        T_c^2 &=\frac{\mu_s^2}{2a_s} +\frac{u_3^2}{8a_s\lambda_s}.
    \end{aligned}
\end{equation}
Local stability of this minimum in the Higgs direction additionally requires
\begin{equation}
    -\mu_H^2+2a_hT_c^2+\frac{\lambda_{sh}}{2}s_c^2>0.
\end{equation}

For $s=0$, the finite-temperature potential reduces to
\begin{equation}
V_T(h,0;T)
=
\left(a_hT^2-\frac{\mu_H^2}{2}\right)h^2
+\frac{\lambda_H}{4}h^4.
\end{equation}
For $\lambda_H>0$, the stationarity and degeneracy conditions with respect to the origin admit no solution with $h_c>0$. Thus, within the thermal approximation adopted here, a direct first-order transition from $(0,0)$ to a phase with $h\neq0$ and $s=0$ is absent. This does not exclude a thermal history in which a continuous electroweak transition precedes a subsequent transition involving the singlet.

The possible symmetry-breaking sequences can be classified as
\begin{equation}
\begin{aligned}
&O(0,0)\longrightarrow B(h,s),
\\
&O(0,0)\longrightarrow A(0,s)\longrightarrow B(h,s),
\\
&O(0,0)\longrightarrow C(h,0)\longrightarrow B(h,s).
\end{aligned}
\end{equation}
The first sequence is a one-step transition, while the second and third proceed through singlet-only and Higgs-only intermediate phases, respectively. In this work, we focus on the first two sequences. For each first-order stage, the critical temperature is determined by the degeneracy of the two relevant minima separated by a potential barrier.

For the two-step thermal history, denoted as scenario B, the first transition proceeds from the symmetric phase to a singlet-only phase. The field values at the critical and nucleation temperatures satisfy
\begin{equation}
    \begin{aligned}
        h_c &=0 , \quad s_c>0,\\
        h_n &=0 , \quad s_n>0.
    \end{aligned}
\end{equation}
Electroweak symmetry remains unbroken during this first transition and is broken at a later stage. Since $v(T_n)=h_n=0$, the broken-phase baryon-number preservation criterion introduced above does not apply to this singlet transition. This observation alone does not exclude electroweak baryogenesis during the subsequent electroweak transition, whose viability requires a separate analysis. In scenario B, we focus on the GW signals associated with the singlet transition and domain-wall annihilation and assess their detectability, without imposing the electroweak baryon-number preservation criterion on the singlet transition.

Scenario A(a) requires a mixed minimum with $h_c>0$ and $s_c>0$ to be degenerate with the symmetric minimum at $T_c$. Candidate points must satisfy the stationarity, degeneracy, and local-stability conditions at the critical temperature, as well as the subsequent nucleation requirement and the theoretical and experimental constraints imposed in this study.

Scenario A(b) requires a singlet-only minimum with $h_c=0$ and $s_c>0$ to be degenerate with the symmetric minimum at $T_c$ and $h_n>0$ and $s_n>0$ at $T_n$. Candidate points must satisfy the stationarity, degeneracy, and local-stability conditions at the critical temperature, as well as the subsequent nucleation requirement and the theoretical and experimental constraints imposed in this study.

\subsection{GWs from phase transitions}
We use CosmoTransitions~2.0.2~\cite{Wainwright:2011kj} to compute the three-dimensional bounce action $S_3(T)$. Two key parameters characterizing the phase transition are its strength $\alpha$ and the dimensionless inverse duration $\beta/H_n$:
\begin{equation}
    \alpha =\frac{\Delta\rho}{\rho_R},\quad \frac{\beta}{H_n} =\left.T\frac{d}{dT}\left(\frac{S_3(T)}{T}\right)\right|_{T=T_n}.
\end{equation}
Here, $H_n=H(T_n)$ is the Hubble expansion rate at the nucleation temperature, and
\begin{equation}
\rho_R=\frac{\pi^2}{30}g_*T_*^4
\end{equation}
is the radiation energy density, with $g_*$ denoting the effective number of relativistic degrees of freedom. We take $T_*\simeq T_n$. The parameter $\beta$ characterizes the inverse time scale of the phase transition.

For the GW signal from the phase transition, we include contributions from sound waves and MHD turbulence~\cite{Caprini:2015zlo}:
\begin{equation}
    \Omega_{\mathrm{GW}}^{\mathrm{pt}}(f)h^2
    \simeq\Omega_{\mathrm{sw}}(f)h^2+\Omega_{\text{turb}}(f)h^2.
\end{equation}
Adding the contribution from domain-wall annihilation, the total spectrum is
\begin{equation}
    \Omega_{\mathrm{GW}}(f)h^2 \simeq \Omega_{\mathrm{GW}}^{\mathrm{pt}}(f)h^2+\Omega_{\mathrm{GW}}^{\mathrm{dw}}(f)h^2,
\end{equation}
where $\Omega_{\mathrm{GW}}^{\mathrm{pt}}$ and $\Omega_{\mathrm{GW}}^{\mathrm{dw}}$ denote the phase-transition and domain wall contributions, respectively. Here, $h$ is the reduced present-day Hubble parameter, defined by $H_0=100h\,\mathrm{km\,s^{-1}\,Mpc^{-1}}$. The GW from the domain wall will be introduced in the next section \ref{sectiond}. Apart from $\alpha$ and $\beta$, we should calculate the bubble wall velocity $v_b$, at which the bubble wall expands outward \cite{Steinhardt:1981ct,Alves:2019igs,Bian:2019bsn,Alves:2018jsw},
\begin{align}
        v_b & =\frac{1/\sqrt{3}+\sqrt{\alpha^2 +2\alpha /3}}{1+\alpha}\ .
\end{align}
The present-day peak frequency of the GW spectrum generated by sound waves is given by~\cite{Hindmarsh:2015qta, Hindmarsh:2013xza, Caprini:2015zlo}
\begin{equation}
    f_{\mathrm{sw}} = 1.9\times10^{-5}\,\mathrm{Hz}\,\frac{\beta}{H_*}\frac{1}{v_b}                                \left(\frac{T_*}{100\,\mathrm{GeV}}\right) \left(\frac{g_*}{100}\right)^{1/6},
\end{equation}
where $H_*=H(T_*)\simeq H_n$ under the approximation $T_*\simeq T_n$. Adopting the long-lived acoustic-source approximation, the corresponding present-day GW energy-density spectrum is
\begin{equation}
    \begin{aligned}
        \Omega_{\mathrm{sw}}(f)h^2={}
                                    &2.65\times10^{-6}\left(\frac{\beta}{H_*}\right)^{-1} \left(\frac{\kappa\alpha}{1+\alpha}\right)^2 \left(\frac{g_*}{100}\right)^{-1/3} \\
                                    &\times v_b\left(\frac{f}{f_{\mathrm{sw}}}\right)^3 \left[ \frac{7}{4+3(f/f_{\mathrm{sw}})^2} \right]^{7/2}.
    \end{aligned}
\end{equation}
Here, $\kappa$ is the efficiency factor describing the fraction of the released energy converted into bulk kinetic energy of the plasma, evaluated using the hydrodynamic analysis of Ref.~\cite{Espinosa:2010hh}. This efficiency factor is distinct from the sphaleron fluctuation determinant denoted by the same symbol above. 
Then the peak frequency of the GW from the MHD turbulence in the plasma is given by \cite{Caprini:2009yp}
\begin{equation}
		f_{\text{turb}}=2.7 \times 10^{-5}\frac{\beta}{H}\frac{1}{v_b}\frac{T_*}{100\, \mathrm{GeV}}\left(\frac{g_*}{100}\right)^{1/6}\mathrm{Hz}\ .
\end{equation}
The corresponding GW energy-density spectrum is
\begin{equation}
        \begin{aligned}
            \Omega_{\text{turb}}h^2(f)= & 3.35 \times 10^{-4}\left(\frac{\beta}{H}\right)^{-1}\left(\frac{\epsilon \kappa \alpha}{1+\alpha}\right)^{3/2}\left(\frac{g_*}{100}\right)^{-1/3}v_b \\
            &\quad \frac{(f/f_{\text{turb}})^3(1+f/f_{\text{turb}})^{-11/3}}{1+8\pi fa_0 /(a_* H_*)}\ .
        \end{aligned}
\end{equation}
where the efficiency factor $\epsilon \approx 0.05$, and the Hubble parameter is given by
\begin{equation}
		h_* =a_*H_* = (1.65\times 10^{-5}\mathrm{Hz})(\frac{T_*}{100\,\mathrm{GeV}})(\frac{g_*}{100})^{1/6}\ .
\end{equation}

\subsection{GWs from domain-wall annihilation}\label{sectiond}

Domain walls can form when the spontaneous breaking of the $\mathbb{Z}_3$ symmetry causes different spatial regions to settle into distinct degenerate vacua~\cite{Durrer:2001cg}. Their formation is therefore associated with singlet symmetry breaking and need not occur after electroweak symmetry breaking. To estimate the domain-wall properties, we adopt a zero-temperature, fixed-modulus approximation, neglecting variations of the singlet modulus and the Higgs background across the wall. We set $h=v$, $\Phi=0$, and parametrize the singlet as~\cite{Zhou:2020ojf}
\begin{equation}
S=\frac{v_s}{\sqrt{2}}e^{i\phi}.
\end{equation}
The tree-level potential then becomes
\begin{equation}
\begin{aligned}
V_0(\phi)={}&
-\frac{\mu_H^2}{2}v^2
-\frac{\mu_s^2}{2}v_s^2
+\frac{\lambda_H}{4}v^4
+\frac{\lambda_s}{4}v_s^4
\\
&+\frac{\lambda_{sh}}{4}v^2v_s^2
+\frac{u_3v_s^3}{2\sqrt{2}}\cos(3\phi).
\end{aligned}
\end{equation}
Substituting this parametrization into the singlet kinetic term gives
\begin{equation}
\mathcal{L}_{\mathrm{kin}}
=
(\partial_\mu S^\dagger)(\partial^\mu S)
=
\frac{v_s^2}{2}(\partial_\mu\phi)(\partial^\mu\phi)
=
\eta^2(\partial_\mu\phi)(\partial^\mu\phi),
\end{equation}
where $\eta^2=v_s^2/2$.

The phase dynamics follow from
\begin{equation}
\mathcal{L}_\phi
=
\frac{v_s^2}{2}(\partial_\mu\phi)(\partial^\mu\phi)
-V_0(\phi).
\end{equation}
For a static planar wall with a profile depending only on $z$, the equation of motion is
\begin{equation}
v_s^2\frac{d^2\phi}{dz^2}
=
\frac{\partial V_0}{\partial\phi},
\end{equation}
or equivalently,
\begin{equation}
\frac{d^2\phi}{dz^2}
+\frac{3u_3v_s}{2\sqrt{2}}\sin(3\phi)=0.
\end{equation}
For $u_3<0$ and $v_s>0$, a solution interpolating between the adjacent vacua $\phi=0$ and $\phi=2\pi/3$ is
\begin{equation}
\phi(z)
=
\frac{4}{3}\arctan\!\left[e^{m(z-z_0)}\right],
\quad
m^2=\frac{9|u_3|v_s}{2\sqrt{2}}=m_\chi^2,
\end{equation}
where $z_0$ is the wall center and $m^{-1}$ sets the characteristic wall thickness.

The domain wall is orthogonal to the z-axis \cite{Hattori:2015xla}. In this assumption, the domain wall tension is estimated as,
	\begin{equation}
        \begin{aligned}
            \sigma_{\text{wall}}&= \int dz \rho_{\text{wall}}(z)\\
            &=\int \left(\left|\frac{dS}{dz}\right|^2+V(S(z),\frac{v}{\sqrt{2}})-V(\frac{v_s}{\sqrt{2}},\frac{v}{\sqrt{2}})\right)dz\ .
        \end{aligned}
	\end{equation}

As the domain wall annihilates, gravitational wave signals will be generated. The peak frequency is given by \cite{Hiramatsu:2013qaa}
	\begin{equation}
    \begin{aligned}
        f^{\text{dw}}_{\text{peak}} \simeq&\ 3.99\times 10^{-9}\ \mathrm{Hz}\ \mathcal{A}^{-1/2} \\
        & \times \left(\frac{1\mathrm{TeV}^3} {\sigma_{\text{wall}}}\right)^{1/2}\left(\frac{\Delta V}{1\ \mathrm{MeV}^4}\right)^{1/2}\ ,\\
        \hat{\sigma}_{\text{wall}}=&\ \frac{\sigma_{\text{wall}}}{1\ \mathrm{TeV}^3},\quad \Delta \hat{V}=\frac{\Delta V}{1\ \mathrm{MeV}^4}
    \end{aligned}
	\end{equation}
The existence of $\Delta V$ breaks the $\mathbb{Z}_3$ symmetry, which determines the decay time of the domain wall. 
Assuming that domain walls annihilate during radiation domination, the present-day peak amplitude of the resulting GW spectrum is estimated as~\cite{Hiramatsu:2013qaa, Kadota:2015dza}
\begin{equation}
\begin{aligned}
\Omega_{\mathrm{GW}}^{\mathrm{dw}}(f_{\mathrm{peak}}^{\mathrm{dw}})h^2
\simeq {} & 5.20\times10^{-20}\,
\tilde{\epsilon}_{\mathrm{GW}}\mathcal{A}^{4}
\left(\frac{10.75}{g_*}\right)^{1/3} \\
& \times
\left(\frac{\sigma_{\mathrm{wall}}}{1\,\mathrm{TeV}^{3}}\right)^{4}
\left(\frac{1\,\mathrm{MeV}^{4}}{\Delta V}\right)^{2},
\end{aligned}
\end{equation}
where $\tilde{\epsilon}_{\mathrm{GW}}$ is the efficiency parameter for GW emission and $g_*$ is the effective number of relativistic degrees of freedom at domain-wall annihilation.

Domain walls must annihilate before they dominate the energy density of the Universe, which imposes the following constraint:
\begin{equation}
		\begin{aligned}
			&\sigma_{\text{wall}} < 2.93\times 10^4 \ \mathrm{TeV}^3\mathcal{A}^{-1}\left(\frac{0.1sec}{t_{dec}}\right)\ ,\\
			&t_{dec}\approx \mathcal{A}\sigma_{\text{wall}}/(\Delta V)\ .
		\end{aligned}
\end{equation}
To preserve the successful predictions of Big Bang nucleosynthesis (BBN), domain walls must annihilate before BBN, which imposes the following constraints~\cite{Kawasaki:2004yh, Kawasaki:2004qu}:
\begin{equation}
		\begin{aligned}
			&t_{dec} \le 0.01\ \text{sec}\ ,\\
			&\Delta V \gtrsim 6.6\times 10^{-2}\ \mathrm{MeV}^4\mathcal{A}\left(\frac{\sigma_{wall}}{1\ \mathrm{TeV}^3}\right)\ .
		\end{aligned}
\end{equation}
In this study, we adopt $\mathcal{A}=1.2$ for the $\mathbb{Z}_3$ symmetry~\cite{Kadota:2015dza}, set the GW emission efficiency to $\tilde{\epsilon}_{\text{GW}}=0.7$~\cite{Kadota:2015dza}, and take the effective number of relativistic degrees of freedom at domain-wall annihilation to be $g_*=10.75$~\cite{Kadota:2015dza, Hiramatsu:2010yz}. The spectral shape is approximated by
\begin{equation}
		\begin{aligned}
			&\Omega^{\text{dw}}_{\text{GW}} h^2(f<f^{\text{dw}}_{\text{peak}})=\Omega^{\text{dw}}_{\text{GW}}h^2_{\text{peak}}\left(\frac{f}{f^{\text{dw}}_{\text{peak}}}\right)^3\ ,\\
			&\Omega^{\text{dw}}_{\text{GW}} h^2(f>f^{\text{dw}}_{\text{peak}})=\Omega^{\text{dw}}_{\text{GW}}h^2_{\text{peak}}\left(\frac{f^{\text{dw}}_{\text{peak}}}{f}\right)\ .
		\end{aligned}
\end{equation}
We fix the present-day peak frequency at $f_{\mathrm{peak}}^{\mathrm{dw}}\simeq2\times10^{-9}\,\mathrm{Hz}$, in the nanohertz band probed by the Parkes Pulsar Timing Array (PPTA). For each allowed parameter point, we calculate the domain-wall tension $\sigma_{\mathrm{wall}}$ and determine the corresponding energy-density bias $\Delta V$ from the peak-frequency relation given above.

\section{Parameter scan and discussion}\label{parameter_scan}

\subsection{Parameter-space scan}
We implement the model in SARAH-4.15.2~\cite{Staub:2015kfa} and use SPheno-4.0.7~\cite{Porod:2011nf, Porod:2003um} in our parameter scan. We take $m_2$, $m_{\chi}$, $m_s$, $v_s$, $\lambda$, $\lambda_1$, $\lambda_2$, $\lambda_3$, and $\theta$ as input parameters and scan over the following ranges:
\begin{equation}
    \begin{aligned}
        &m_{\chi}\in[200,6000]\,\mathrm{GeV},
        \qquad m_2\in[150,3000]\,\mathrm{GeV},\\
        &m_s\in[300,1000]\,\mathrm{GeV},
        \qquad v_s\in[150,10000]\,\mathrm{GeV},\\
        &\lambda_1\in[0,12],\qquad \lambda_3\in[-12,12],\qquad \lambda\in[0,12],\\
        &\sin\theta\in[0,0.37].
    \end{aligned}
\end{equation}
The coupling $\lambda_2$ is chosen to satisfy
\begin{equation}
    |\lambda_2+2\lambda_3|<0.1
    \quad\text{or}\quad
    |\lambda_2-2\lambda_3|<0.1.
\end{equation}
We further impose the following theoretical and experimental constraints.
\begin{itemize}
    \item [1.]Tree-level boundedness from below. With $\lambda_4=\cdots=\lambda_{11}=0$ and $m_s^2>0$, we impose the following sufficient conditions to ensure that the tree-level scalar potential is bounded from below:
    \begin{equation}
        \quad\begin{aligned}
            &\lambda_H>0,\qquad \lambda_s>0,\qquad \lambda_{sh}+2\sqrt{\lambda_H\lambda_s}>0,\\
            &\lambda>0,\qquad \lambda_1>0,\\
            &\lambda_1+\lambda_2+2\lambda_3>0,\qquad \lambda_1+\lambda_2-2\lambda_3>0.
        \end{aligned}
    \end{equation}
    \item [2.]Perturbativity and tree-level unitarity. We impose the following perturbativity requirements on the scalar quartic couplings:
    \begin{equation}
        \quad\begin{aligned}
            &|\lambda_H|,\ |\lambda_s|,\ |\lambda_{sh}|,\ |\lambda|,\ 
            |\lambda_1|,\ |\lambda_2|,\ |\lambda_3|<4\pi,\\
            &|\lambda_1+\lambda_2+2\lambda_3|\leq4\pi,\\
            &|\lambda_1+\lambda_2-2\lambda_3|\leq4\pi.
        \end{aligned}
    \end{equation}
    For high-energy scalar scattering involving the Higgs doublet and the complex singlet, we additionally impose the tree-level unitarity condition
    \begin{equation}
        \quad \left|3\lambda_H+2\lambda_s
        \pm\sqrt{(3\lambda_H-2\lambda_s)^2+2\lambda_{sh}^{2}}\right|<8\pi.
    \end{equation}
    \item [3.]Constraints from the observed Higgs boson. We identify $h_1$ with the observed Higgs boson of mass approximately $125\,\mathrm{GeV}$. We use HiggsSignals within HiggsTools~\cite{Bahl:2022igd} to evaluate $\chi^2_{125}$ by comparing the model predictions with the LHC Higgs measurements included in the dataset used in our analysis. We retain parameter points satisfying
    \begin{equation}
        \quad \chi^2_{125}-\chi^2_{\mathrm{SM}}<6.18,
    \end{equation}
    where $\chi^2_{\mathrm{SM}}$ is evaluated for the SM using the same dataset and statistical settings. We adopt this condition as a selection criterion relative to the SM fit.
    \item [4.]Electroweak precision constraints. We calculate the oblique parameters $S$ and $T$ using SPheno-4.0.7~\cite{Porod:2011nf,Porod:2003um} and compare them with the fit results obtained assuming $U=0$~\cite{ParticleDataGroup:2026aaa},
    \begin{equation}
        \quad S=0.008\pm0.071,\qquad T=0.021\pm0.055,
    \end{equation}
    with a correlation coefficient of $0.92$. Including this correlation, we require $\chi^2_{ST}<6.18$, corresponding to the joint $2\sigma$ confidence region for two degrees of freedom.
\end{itemize}

\subsection{Results and discussion}

\begin{figure*}[htbp]
		\centering
		\includegraphics[width=0.95\textwidth]{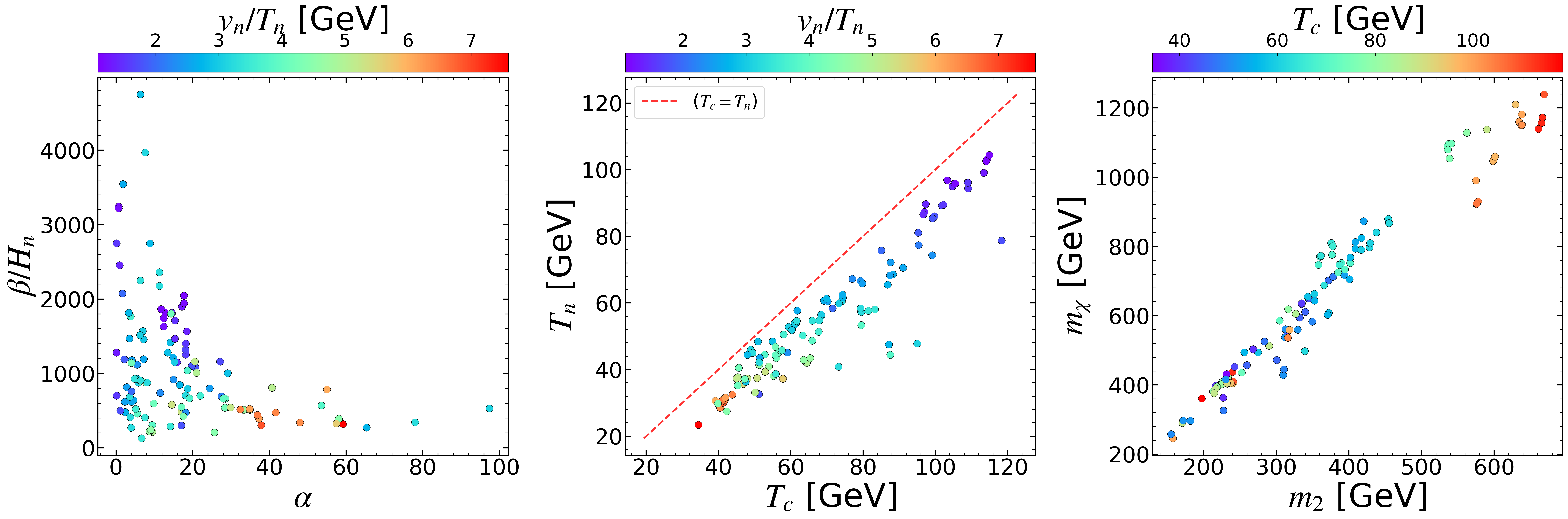}
		\caption{In scenario A, left: The relations between $\alpha$ and $\beta/H_n$ with the nucleation temperature. The horizontal axis represents $\alpha$, and the vertical axis represents $\beta/H_n$. Furthermore, the color bar represents $T_n$; Center: The relations between nucleation temperature and the critical temperature with the strong first-order parameter. The horizontal axis represents $T_c$, and the vertical axis represents $T_n$. Furthermore, the color bar represents $v_n/T_n$. Right: The relations between $m_2$ and $m_{\chi}$ with $T_c$. The horizontal axis represents $m_2$, and the vertical axis represents $m_{\chi}$. The color bar represents $T_c$.}
		\label{fig2}
\end{figure*}
\begin{figure*}[htbp]
		\centering
		\includegraphics[width=1\textwidth]{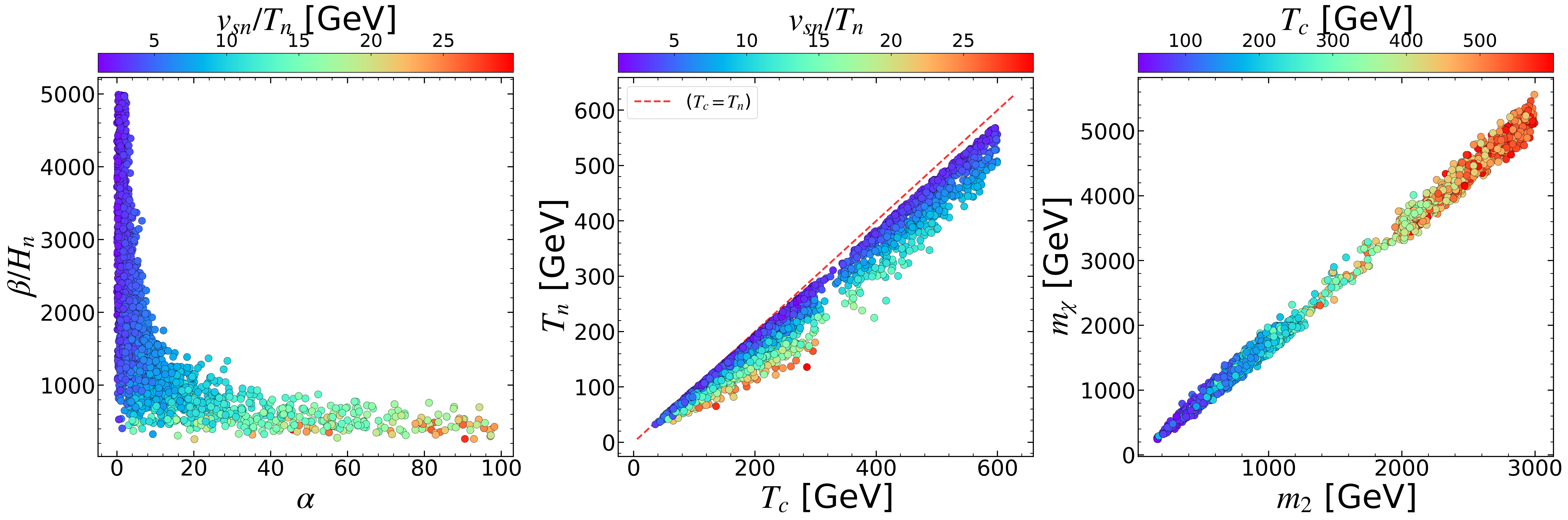}
		\caption{Same as Fig.~\ref{fig2}, but for scenario B. The color in the center panel indicates the singlet transition-strength ratio $v_{sn}/T_n$. All transition quantities refer to the first transition along the singlet direction.}
		\label{fig3}
\end{figure*}
\begin{figure*}[htbp]
\centering
\includegraphics[width=\textwidth]{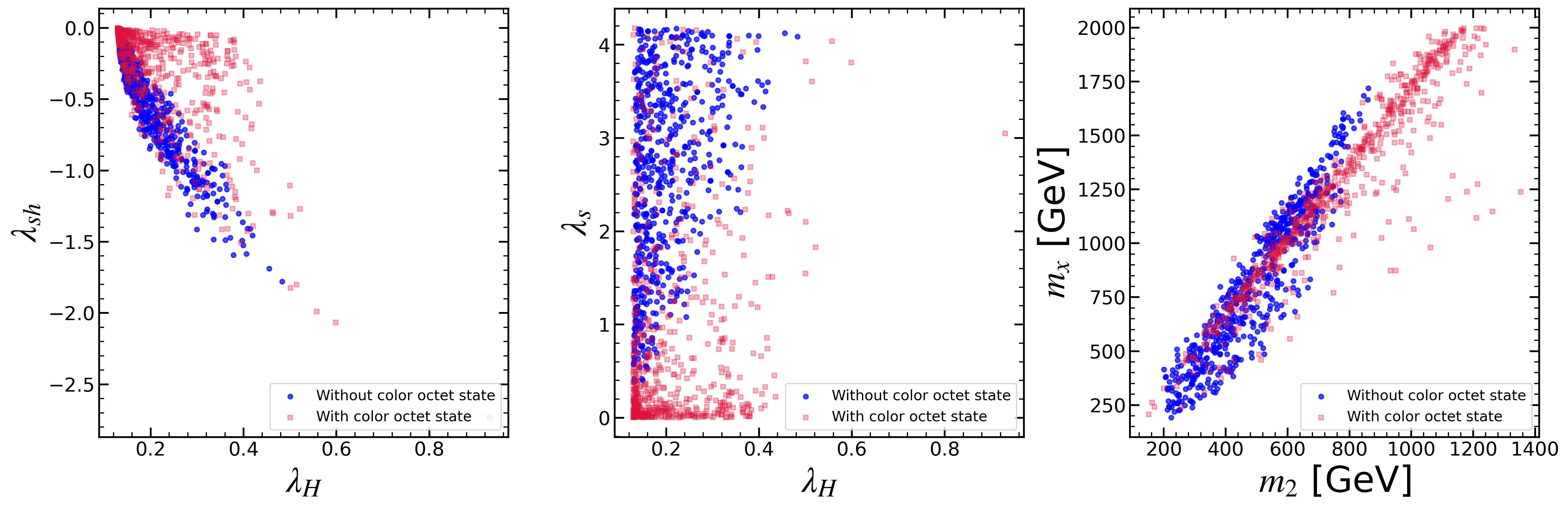}
\caption{Comparison between the MWS model and the complex-singlet-extended SM in the $(\lambda_H,\lambda_{sh})$ plane (left), the $(\lambda_H,\lambda_s)$ plane (center), and the $(m_2,m_\chi)$ plane (right). Red squares represent the MWS model, while blue circles represent the complex-singlet-extended SM.}
\label{fig4}
\end{figure*}

Figures~\ref{fig2} and~\ref{fig3} show the distributions of the selected parameter points in scenarios A and B, respectively. In each figure, the left panel displays $\beta/H_n$ as a function of $\alpha$, with the color indicating the nucleation temperature $v_{sn}/T_n$. The middle panel shows $T_n$ versus the critical temperature $T_c$, with the color representing $v_n/T_n$ in scenario A and $v_{sn}/T_n$ in scenario B. The red dashed line denotes $T_n=T_c$. The right panel displays $m_{\chi}$ versus $m_2$, with the color indicating $T_c$. In all panels, the color scale runs from purple for smaller values to yellow for larger values. For scenario B, the transition quantities refer to the first transition along the singlet direction.

In scenario A, the left panel of Fig.~\ref{fig2} shows that points with large $\alpha$ generally have relatively large $v_{sn}/T_n$, whereas points with large $\beta/H_n$ are concentrated at small $\alpha$. This suggests that stronger supercooling can favor a larger transition strength, although the correlations are not strictly monotonic. In the middle panel, all points lie below the line $T_n=T_c$, and larger values of $v_n/T_n$ occur predominantly at lower $T_n$. The broad distribution below the diagonal indicates that the critical temperature alone does not determine the nucleation temperature or the transition strength. The right panel exhibits a positive correlation between $m_2$ and $m_{\chi}$, with larger masses tending to be associated with higher critical temperatures.

Scenario B exhibits similar overall trends, but with a broader range of transition temperatures, as shown in Fig.~\ref{fig3}. Large values of $\beta/H_n$ occur mainly at small $\alpha$, while the large-$\alpha$ region is predominantly populated by points with relatively large $v_{sn}/T_n$. Nevertheless, the substantial scatter prevents a one-to-one correspondence between $T_n$ and either $\alpha$ or $\beta/H_n$. The middle panel shows a positive correlation between $T_c$ and $T_n$. At comparable $T_c$, points with lower $T_n$ generally have larger $v_{sn}/T_n$, indicating an association between stronger supercooling and a larger singlet order parameter normalized by temperature. The right panel again shows a positive correlation between $m_2$ and $m_{\chi}$, and the heavier-mass region tends to have larger $T_c$. Since this first transition leaves electroweak symmetry unbroken, $v_{sn}/T_n$ characterizes the singlet transition rather than baryon-number preservation.

Figure~\ref{fig4} compares the sampled parameter points in the MWS model with those in the complex-singlet extension of the SM~\cite{Zhou:2020ojf}. The left and middle panels show $\lambda_{sh}$ and $\lambda_s$, respectively, as functions of $\lambda_H$, while the right panel displays $m_{\chi}$ versus $m_2$. Red squares represent the MWS model, and blue circles represent the model without color-octet scalars. The two samples partially overlap, but the MWS sample extends to larger values of $\lambda_H$, $m_2$, and $m_{\chi}$ and contains more points with small $\lambda_s$. The additional color-octet interactions modify the thermal corrections to the Higgs and singlet potentials and can therefore alter the phase-transition dynamics. These distributions illustrate differences between the selected samples; they do not establish the full boundaries of the allowed parameter spaces.

\begin{figure}[htbp]
		\centering
		\includegraphics[width=0.5\textwidth]{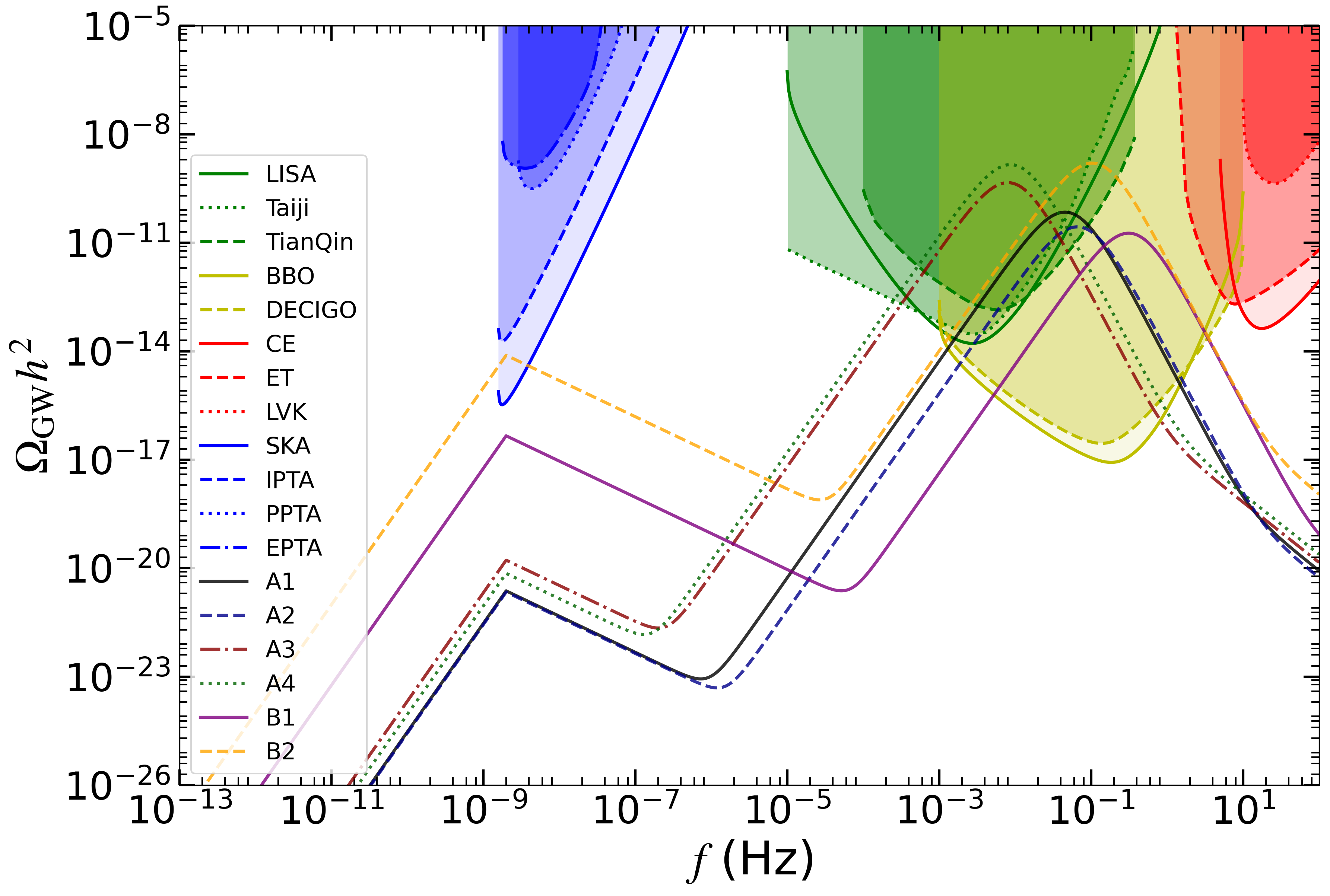}
		\caption{Gravitational-wave spectra from first-order phase transitions and domain-wall annihilation in scenarios A and B. The phase-transition contribution arises from the one-step electroweak transition in scenario A and the first transition along the singlet direction in scenario B. Observational limits and projected sensitivity curves are shown for LISA \cite{LISA:2017pwj}, DECIGO \cite{Kudoh:2005as}, BBO \cite{Corbin:2005ny}, TianQin \cite{TianQin:2015yph}, Taiji \cite{Gong:2021any}, EPTA \cite{Antoniadis:2023ept1, Antoniadis:2023ept2}, PPTA \cite{Reardon:2023ppta}, IPTA \cite{Verbiest:2016vem}, SKA \cite{Lazio_2013}, CE \cite{Shoemaker:2026rfx, Reitze2019}, ET \cite{ET:2019dnz}, and LVK \cite{KAGRA:2013rdx, LIGOScientific:2016vlm}.}
		\label{fig5}
	\end{figure}

Figure~\ref{fig5} presents the GW spectra for the six benchmark points listed in Table~\ref{tab:benchmark-points}. The horizontal axis is the present-day frequency $f$, and the vertical axis is the present-day GW energy density $\Omega_{\mathrm{GW}}h^2$, both on logarithmic scales. The black solid, blue dashed, brown dash-dotted, and green dotted curves correspond to $\mathrm{A_1}$, $\mathrm{A_2}$, $\mathrm{A_3}$, and $\mathrm{A_4}$, respectively. The purple solid and orange dashed curves correspond to $\mathrm{B_1}$ and $\mathrm{B_2}$. The first two benchmarks belong to scenario A(a), the next two to scenario A(b), and the last two to scenario B. The remaining curves and shaded regions show the observational limits and projected sensitivities of the GW experiments indicated in the legend.

The benchmark spectra exhibit two well-separated peaks. The high-frequency peak, at approximately $10^{-3}$--$10^{-2}\,\mathrm{Hz}$, arises from the first-order phase transition, whereas the low-frequency peak originates from domain-wall annihilation. In scenario A, the phase-transition contribution is generated by the one-step electroweak transition. In scenario B, it is generated by the first, singlet-direction transition; a possible contribution from the subsequent electroweak transition is not included here. The low-frequency peak is fixed at approximately $2\times10^{-9}\,\mathrm{Hz}$ by our choice of $\Delta V$ for each parameter point, rather than being an independent prediction of the scan.

For the phase-transition contribution, the peak frequency depends on $T_n$, $\beta/H_n$, and the bubble-wall velocity, while the amplitude depends on the transition strength, the duration, and the efficiency of energy transfer to the plasma. A smaller $\beta/H_n$ generally enhances the signal when the other quantities are held fixed. However, a larger $\alpha$ does not imply an unlimited increase in the amplitude, since the factor $\alpha/(1+\alpha)$ approaches unity for $\alpha\gg1$. The spectra shown overlap the sensitivity regions of space-based observatories such as LISA~\cite{LISA:2017pwj}, DECIGO~\cite{Kudoh:2005as}, BBO~\cite{Corbin:2005ny}, TianQin~\cite{TianQin:2015yph}, and Taiji~\cite{Gong:2021any}. This comparison suggests potential observability within the adopted GW prescription, although a quantitative detection forecast requires a signal-to-noise analysis.

The domain-wall contribution provides a more pronounced distinction between the two scenarios. In the scenario A sample, the predicted signals remain below the PTA sensitivity curves considered here, including the projected SKA sensitivity. The selected scenario A benchmarks illustrate relatively large domain-wall amplitudes within this scan, but their low-frequency peaks are still too weak to reach these curves. By contrast, the domain-wall signals of $\mathrm{B_1}$ and $\mathrm{B_2}$ extend above the projected SKA sensitivity over part of the nanohertz band, while remaining below the EPTA, PPTA, and IPTA curves shown.

This difference can be understood from the domain-wall tension. For fixed $\mathcal{A}$ and fixed $f_{\mathrm{peak}}^{\mathrm{dw}}$, the peak-frequency relation implies $\Delta V\propto\sigma_{\mathrm{wall}}$. Consequently, with the remaining quantities held fixed, the peak amplitude scales as
\begin{equation}
\Omega_{\mathrm{GW}}^{\mathrm{dw}}(f_{\mathrm{peak}}^{\mathrm{dw}})h^2
\propto\frac{\sigma_{\mathrm{wall}}^4}{(\Delta V)^2}
\propto\sigma_{\mathrm{wall}}^2.
\end{equation}
The scenario B benchmarks have substantially larger domain-wall tensions than the scenario A benchmarks, accounting for their stronger low-frequency signals. Their large singlet VEVs contribute to this enhancement, although $v_s$ alone does not determine the tension or the GW amplitude.

% del 
Several benchmarks have $\alpha\gg1$, so the phase-transition spectra should be interpreted as estimates within the approximations adopted here. 
Reliable predictions in this regime require checking transition completion, the expansion history and reheating, and the lifetime of the acoustic source. 
Subject to these qualifications, the results illustrate the complementary roles of space-based interferometers and PTAs: the former probe the phase-transition contribution, while the latter can test domain-wall annihilation in parameter regions with sufficiently large wall tensions.

\begin{table}[htbp]
    \centering
    \renewcommand{\arraystretch}{1.15}
    \caption{Key parameters of the benchmark points shown in Fig.~\ref{fig5}.}
    \begingroup
    \begin{tabular}{ccccccc}
        \hline
        \hline
        Parameter
        & $\mathrm{A_1}$
        & $\mathrm{A_2}$
        & $\mathrm{A_3}$
        & $\mathrm{A_4}$
        & $\mathrm{B_1}$
        & $\mathrm{B_2}$ \\
        \hline
        $m_2$ [GeV]
        & 318.36 & 316.28 & 197.66 & 304.84 & 1984.12 & 2626.04 \\
        $\theta$
        & 0.10 & 0.09 & 0.18 & 0.01 & 0.05 & 0.00 \\
        $v_s$ [GeV]
        & 327.90 & 328.65 & 619.10 & 393.69 & 1518.06 & 4932.11 \\
        $m_s$ [GeV]
        & 544.21 & 536.66 & 637.99 & 454.41 & 500.89 & 451.26 \\
        $\lambda$
        & 2.52 & 2.35 & 0.69 & 4.70 & 3.97 & 6.42 \\
        $\lambda_1$
        & 0.12 & 0.06 & 0.62 & 1.37 & 2.14 & 3.77 \\
        $\lambda_2$
        & 0.04 & 0.004 & -0.17 & 1.38 & 0.55 & 0.10 \\
        $\lambda_3$
        & -0.02 & -0.002 & 0.11 & 0.70 & 0.30 & 0.04 \\
        $m_{\chi}$ [GeV]
        & 558.73 & 535.86 & 360.50 & 585.37 & 3629.59 & 4472.14 \\
        $T_n$ [GeV]
        & 89.62 & 96.79 & 78.68 & 58.33 & 465.85 & 505.28 \\ 
        $\alpha$
        & 0.96 & 0.71 & 1.13 & 4.06 & 0.59 & 8.86 \\ 
        $\beta/H_n$ 
        & 2454.24 & 3242.30 & 498.69 & 758.61 & 3153.94 & 1055.92 \\ 
        $\hat{\sigma}_{wall}$
        & 0.05 & 0.05 & 0.14 & 0.09 & 7.44 & 96.70 \\ 
        $\Delta \hat{V}$
        & 0.016 & 0.016 & 0.043 & 0.028 & 2.24 & 29.16 \\ \hline\hline
        
    \end{tabular}
    \label{tab:benchmark-points}
    \endgroup
\end{table}

\section{\label{sec:conclusion}Conclusions}

We have studied first-order phase transitions and the associated GW signals in the Manohar--Wise model extended by a complex scalar singlet. The spontaneous breaking of the $\mathbb{Z}_3$ symmetry produces domain walls, whose subsequent annihilation is induced by a small explicit symmetry-breaking bias. After applying the theoretical and experimental selection criteria described above, we analyzed the phase-transition histories and estimated the GW contributions from the transitions and domain-wall annihilation.

The selected parameter points exhibit two distinct thermal histories. Scenario A features a one-step, strongly first-order electroweak phase transition that satisfies the BNPC. Scenario B follows a two-step history, with the first transition occurring along the singlet direction while electroweak symmetry remains unbroken. Our GW analysis in scenario B concerns this first transition, so its implications for electroweak baryogenesis require a separate study of the subsequent electroweak transition. Compared with the sample for the complex-singlet extension of the SM, the MWS sample extends to larger values of $\lambda_H$, $m_2$, and $m_{\chi}$ and includes points with smaller $\lambda_s$, illustrating the impact of the additional color-octet interactions on the sampled phase-transition parameter space.

Within the adopted GW approximations, the benchmark spectra display a high-frequency peak from the first-order phase transition and a low-frequency peak from domain-wall annihilation. The phase-transition peaks lie at approximately $10^{-3}$--$10^{-2}\,\mathrm{Hz}$ and have potential sensitivity overlap with space-based GW observatories in both scenarios. The domain-wall peak is fixed at approximately $2\times10^{-9}\,\mathrm{Hz}$ in our analysis. Its amplitude remains below the PTA sensitivity curves considered for the scenario A sample, whereas the larger wall tensions of the scenario B benchmarks yield signals that reach the projected SKA sensitivity. Thus, the domain-wall contribution provides an additional observational handle on the thermal histories represented by these samples.

These results motivate complementary searches across the millihertz and nanohertz frequency bands. A two-peak stochastic background would be consistent with the combination of a first-order phase transition and domain-wall annihilation studied here, although it would not uniquely identify this model. Quantitative detectability forecasts require a dedicated signal-to-noise analysis and, for the strongly supercooled benchmarks, further validation of the transition dynamics and GW production assumptions.
 
\section*{Acknowledgments}

This work was supported by the National Natural Science Foundation of China under Grant No. 12275066 and by the startup research funds of Henan University.

%\newpage
% whether or not they are referenced in the text. This is an appropriate
% \nocite{*}

% \bibliographystyle{apsrev4-2}
\bibliographystyle{apsrev4-1}
\bibliography{apssamp}% Produces the bibliography via BibTeX.

\end{document}